\documentclass[useAMS,usenatbib,twocolumn,usegraphicx,preprint]{mn2e}
\usepackage{amsmath}
\usepackage{graphicx}
\usepackage{oldlfont}
\usepackage{longtable}
\usepackage{rotating}
\usepackage{amssymb}

\newcommand{\beq}{\begin{equation}}
\newcommand{\eeq}{\end{equation}}
\newcommand{\beqar}{\begin{eqnarray}}
\newcommand{\eeqar}{\end{eqnarray}}

\title[On the Quasi-Periodic Oscillations of  Magnetars ] {On the Quasi-Periodic Oscillations of  Magnetars }

\author[ A. Colaiuda, H. Beyer  \& K. D. Kokkotas]
{ A.
Colaiuda$^1$\thanks{E-mail:colaiuda@tat.physik.uni-tuebingen.de},
H. Beyer$^2$\thanks{E-mail:horst@cct.lsu.edu}
and K. D. Kokkotas$^{1,3}$\thanks{E-mail:kostas.kokkotas@uni-tuebingen.de}
\\
  $^1$Theoretical Astrophysics, University of T\"{u}bingen, Auf der Morgenstelle 10, 72076, T\"{u}bingen, Germany\\
  $^2$ Instituto de Fisica y Matematicas, Universidad Michoacana de San Nicolas de Hidalgo Edificio C-3, \\
\phantom{$^2$} Ciudad Universitaria Morelia, Michoacan C.P. 58040, Mexico \\
  $^3$Department of Physics, Aristotle University of Thessaloniki, Thessaloniki 54124, Greece \\}
\begin{document}
\maketitle

\begin{abstract}
We study torsional Alfv\'en oscillations of 
magnetars, i.e., neutron stars 
with a strong magnetic field. We consider the 
poloidal and toroidal components 
of the magnetic field and a wide range of equilibrium 
stellar models. We use 
a new coordinate 
system $(X,Y)$, where  $X=\sqrt{a_1} \sin \theta$, $Y=\sqrt{a_1}\cos \theta$ 
and $a_1$ is the radial component of the magnetic field. 
In this coordinate system,
the $1+2$-dimensional evolution equation describing the quasi-periodic 
oscillations, QPOs, 
see \cite{SKS2007}, is 
reduced to a $1+1$-dimensional equation, where the
perturbations propagate only along the $Y$-axis. 
We solve the $1+1$-dimensional equation for different 
boundary conditions and open magnetic field lines, i.e., 
magnetic field lines that reach the surface and there 
match up with
the exterior dipole magnetic field, as well as closed magnetic 
lines, i.e., magnetic lines that never reach the stellar surface. 
For the open field lines, we find two families of 
QPOs frequencies; a family 
of ``lower'' QPOs frequencies which is located near the $X$-axis 
and a family of ``upper'' frequencies located near the $Y$-axis. 
According to \cite{L2007}, the  fundamental frequencies of these 
two families can be 
interpreted as the turning 
points of a continuous spectrum. 
We find that the upper frequencies are constant multiples of 
the lower frequencies with a constant equaling $2n+1$. For the 
closed lines, the corresponding factor is $n+1$ . By 
these relations, we can explain both
the lower and the higher observed frequencies in 
SGR 1806-20 and SGR 1900+14.    
\end{abstract}
\section{Introduction}
\label{sec:Intro}

Strongly magnetized compact stars, so called `magnetars', reveal their 
presence via giant flares with peak luminosities of $10^{44}-10^{46}$ ergs/s. The giant flares are typically accompanied by 
a decaying tail which may last several hundreds of seconds.  
These sources are known as Soft Gamma Repeaters (SGRs), and 
up to now, three giant flares have been detected, SGR 0526-66
in 1979, SGR 1900+14 in 1998, and SGR 1806-20 in 2004. The analysis 
of the timing of the last two events revealed several QPOs
in the decaying tail whose frequencies are approximately
18, 26, 30, 92, 150, 625, and 1840 Hz for SGR 1806-20,
and 28, 53, 84, and 155 Hz for SGR 1900+14, see \cite{WS2006}.
It is believed that during an SGR event
torsional oscillations in the solid crust of the star 
could be excited, \citep{Duncan1998},
that lead to the observed frequencies in the X-ray tail. Indeed,  
the frequency of many of these oscillations fit the values of 
the torsional mode oscillations of the solid crust of a compact star. 
However, since not all of the observed frequencies are explainable 
by pure crust oscillations, \cite{SKS2007,SA2007},  
alternative scenarios have also been suggested. For example, 
\cite{GSA2006} claimed that the observed 
spectra may be explained via  Alfv\'{e}n
oscillations, while \cite{Levin2006} pointed out that 
the Alfv\'{e}n
oscillations of magnetars could be continuous. Recently, 
a toy model calculation by
\cite{Levin2007} indicated that the edges or turning points of the
continuum could correspond to long-lived QPOs, and according 
to Levin, the results of  
\cite{SKS2008} indicated for realistic magnetar configurations 
that the Alfv\'{e}n oscillations
of magnetars could be continuous which would explain  
the lower observed frequencies. In a more recent paper 
by \cite{SK2009}, it was shown that the spectrum of the 
polar Alfv\'{e}n oscillations is actually discrete.

 In \cite{SKS2008}, the authors performed two-dimensional numerical simulations of 
linearized 
Alfv\'{e}n oscillations in magnetars. Their model improves  the previously 
considered toy models in various ways. General Relativity is assumed, various 
realistic equations of state (EOS) are considered and a consistent dipolar magnetic
field is constructed. However, it does not take into account the presence of a 
solid crust and 
only examines the response of the ideal magnetofluid to a chosen initial perturbation. 
The two-dimensional partial differential equation (PDE) that was used 
to study the Alfv\'{e}n oscillations has a mathematical pathological behavior, as will explained later, that forced them to introduce artificial 
numerical viscosity in an attempt 
to stabilize the numerical evolution.
The numerical results presented
in \cite{SKS2008} are compatible with the observations. For example, they
found two families of QPOs, corresponding to the edges or turning points of
a continuum, with harmonics at near integer multiples.  With this identification,
they could set an upper limit to the dipole magnetic field of $\sim3$ to 
$~7\times 10^{15}$G, and they could limit the models to very stiff 
EOS for values of the magnetic field strength near to its upper limit, or 
moderately stiff for lower values of the
magnetic field. 

In an extension of that paper, \cite{SCK2008}, studied the axisymmetric crust 
torsional modes of magnetars
with poloidal and toroidal magnetic field, where both components are confined in the crust.
The numerical results showed that this magnetic configuration cannot explain the actual observational data of SGRs, and is in agreement with a more recent result 
by \cite{vHL2008}.

After a more elaborate study of the master equation, we discovered 
the origin of the numerical instability, and, moreover, we managed to reduce 
it, by using a coordinate transformation, to a $1+1$-dimensional PDE which 
can be evolved stably without the need of artificial viscosity and also allows 
a semi-analytically estimate of the properties of the continuum and the associated QPO's.

In this paper,
we adopt the units of $c=G=1$, where $c$ and $G$ denote the speed
of light and the gravitational constant, respectively, and the
metric signature is $(-,+,+,+)$.

\section{The equation of the problem}
\label{sec:num}

We consider a spherically symmetric and static star described by the
TOV equations and the line element
\begin{equation}\label{ds}
ds^2=-e^{2\Phi(r)}dt^2+e^{2\Lambda(r)}dr^2+r^2(d\theta^2+\sin^2(\theta)d\phi^2) \, \, .
\end{equation}
The structure of the star is not influenced by the presence of the
magnetic field because the last has an energy $E_m$ which is a 
few orders of magnitude
smaller than the gravitational binding energy $E_g$. Typically, $E_m/E_g
\simeq 10^{-4}(B/10^{16}G)^2$.

Generally, spherical stars show two type of oscillations, {\it spheroidal}
oscillations with polar parity and {\it toroidal} oscillations with axial parity.
The observed QPOs in SGR X-ray tails may originate from toroidal oscillations,
since these can be excited more easily than poloidal
oscillations because they do not involve density variations.

The MHD oscillations of the equilibrium model are described by the
linearized equation of motion and by the magnetic induction equations, 
see \cite{SKS2007}. These two perturbation equations can be converted into a 
$1+2$-dimensional evolution equation for the displacement function
${\cal Y}(t,r,\theta)$ which is related to the contravariant component 
of the perturbed 4-velocity, $\delta u^\phi$ via
\begin{equation}
\delta u^\phi = e^{-\Phi} \partial_t {\cal Y} (t,r,\theta) \, .
\end{equation}
For an analytical derivation, see \cite{SKS2008}. 
The $1+2$-dimensional time-evolution equation for the displacement 
${\cal Y}(t,r,\theta)$ has the following form (\cite{SKS2008}):
\begin{equation}\label{2d}
A_{tt}\frac{\partial^2 {\cal Y}}{\partial t^2}=A_{20}\frac{\partial^2 {\cal Y}}{\partial r^2}
+A_{11}\frac{\partial^2 {\cal Y}}{\partial r \partial \theta}+A_{02}\frac{\partial^2 {\cal Y}}{\partial \theta ^2}
+A_{10}\frac{\partial {\cal Y}}{\partial r}+A_{01}\frac{\partial {\cal Y}}{\partial \theta}
\end{equation}
All the coefficients $A_{tt},A_{20},A_{11},A_{02},A_{10}$ and $A_{01}$ depend 
on the coordinates $r$ and $\theta$, but not on time. They are given by 
\begin{equation}\label{att}
A_{tt}=\biggl[\epsilon+\rho+\frac{a_1^2}{\pi r^4}\cos^2(\theta)+\frac{{a_1}'^{2}}{4\pi r^4}e^{-2\Lambda}\sin^2(\theta)\biggr]e^{-2(\Phi-\Lambda)} \, \, , 
\end{equation}
\begin{eqnarray}
\label{a20}
A_{20}&=&\frac{a_1^2}{\pi r^4}\cos^2(\theta)+\mu \, \, , \\
\label{a11}
A_{11}&=&-\frac{a_1{a_1}{'}}{\pi r^4}\cos(\theta)\sin(\theta) \, \, , \\
\label{a02}
A_{02}&=&\frac{{a_1}'^{2}}{4\pi r^4}\sin^2(\theta)+\frac{\mu}{r^2}e^{2 \Lambda} \, \, ,
\end{eqnarray}
\begin{equation}\label{a10}
A_{10}=(\Phi{ '}-\Lambda{ '})\frac{a_1^2}{\pi r^4}\cos^2(\theta)+
\frac{a_1{a_1}'^{}}{2\pi r^4}\sin^2(\theta)+\biggl[\mu{'}+\mu
\biggl(\frac{4}{r}-\Lambda{'}+\Phi{'}\biggr)\biggr] \, \, ,
\end{equation}
\begin{equation}\label{a01}
A_{01}=\biggl[\frac{a_1}{\pi r^4}\biggl(2\pi j_1-\frac{a_1}{r^2}\biggr)+3\frac{{a_1}'^{2}}{4\pi r^4}\biggr]\sin(\theta)\cos(\theta)+\frac{3}{r^2}\mu \cot(\theta)e^{2\Lambda} \, \, ,
\end{equation}
where $a_1(r)$ is the radial component of the electromagnetic four-potential 
and $j_1(r)$ is the four-current. A prime indicates a derivative with respect  
the radius. The symbol $\mu$ denotes the shear modulus. In the following, we 
will set $\mu$ equal to zero, neglecting the presence of a solid crust, since 
we will focus on the oscillations of the magneto-fluid core
that plays a major role in magnetars QPOs. In fact, as it has been shown by   
\cite{Levin2006} and  \cite{GSA2006}, in the 
presence of a strong magnetic field
pure crustal modes cannot explain the present QPOs frequencies. As a consequence, 
it is necessary to study global modes, i.e., Alfv\'en waves propagating through the 
whole star.

The radial function $a_1$ is determined by solving the Grad-Shavranov equation:
\begin{equation}\label{gf}
a_1{''}e^{-2\Lambda}+\left(\Phi'-\Lambda' \right)e^{-2\Lambda}{a_1}'-\frac{2}{r^2}a_1=-4\pi j_1 \, \, .
\end{equation}
As boundary conditions, we assume regularity at the 
center, $a_1\simeq \alpha_0 r^2$, where $\alpha_0$ is a constant, and 
continuity at the surface with a dipole field outside the star.

\begin{figure}
\begin{center}
\includegraphics[height=8cm,angle=-90]{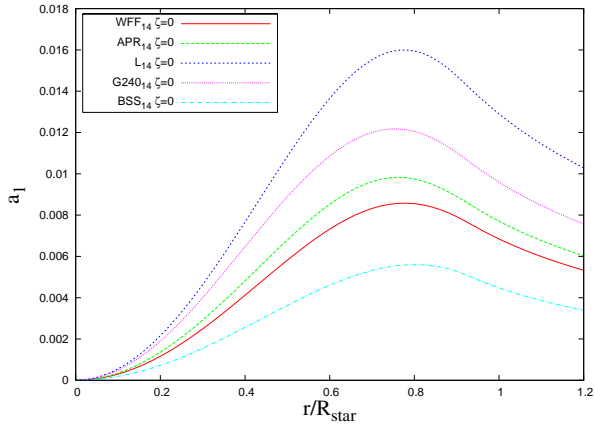}
\caption{
 Plot of $a_1(r)$ for different EOS. The coordinate $r$ is 
normalized to the stellar radius.}%
\label{B_eos}
\end{center}
\end{figure}

According to \cite{SKS2007}, the solutions of equation (\ref{2d}) have to satisfy the  
following boundary conditions:
\begin{itemize}
\item regularity at the center: ${\cal Y}=0$ at $r=0$;
\item no traction on the surface: ${\cal Y}_{,r}=0$ at $r=R$;
\item axisymmetry:  ${\cal Y}_{,\theta}=0$ at $\theta=0$;
\item equatorial plane symmetry for $\ell=2$ initial data:  ${\cal Y}=0$ at $\theta=\pi/2$;
\item for $\ell=3$ initial data: ${\cal Y}_{,\theta}=0$ at $\theta=\pi/2$.
 \end{itemize}
In \cite{SKS2008}, in order to avoid numerical instability, equation (\ref{2d}) 
was solved by introducing an artificial  
$4$-th order `dissipation' term. Here, instead, we find a coordinate transformation 
that allows us to avoid the use of a dissipation term and, as we will show later, 
to avoid the singular behavior of the time independent form of the equation 
(\ref{2d}). In fact, if, in a first approximation, we neglect the first derivatives in the 
direction of $r$ and $\theta$ in the equation (\ref{2d}), then  only the 
coefficients $A_{20}$, $A_{11}$ and $A_{02}$ remain. Calculating the determinant of 
the principal part of the equation (\ref{2d}), we find that 
\begin{equation}\label{det}
\begin{split}
\mathit{D}&=A_{20}A_{02}-\left(\frac{A_{11}}{2}\right)^2 \\&
=\frac{a_1^2 {a_1}'^{2}}{4\pi^2 r^8}\cos^2(\theta)\sin^2 (\theta)-\frac{a_1^2 {a_1}'^{2}}{4\pi^2 r^8}\cos^2(\theta)\sin^2 (\theta) = 0 \, \, .
\end{split}
\end{equation}
Hence the equation is parabolic in every point of its domain, 
and therefore does not describe propagation of 
2D waves. Significant simplification
is achieved by introduction 
of a new coordinate system given by the transformation
\begin{equation}\label{new_trasf}
X=\sqrt{a_1} \sin \theta \;\;\;\;\; Y=\sqrt{a_1} \cos \theta \, \, . 
\end{equation}
This new coordinates system is a kind of Cartesian coordinate system where, 
instead of the radial coordinate $r$, we use the radial component of 
the magnetic four-potential 
$a_1$. The behavior of $a_1$ inside the star is determined via the 
equation (\ref{gf}). On the surface, we require the continuity of $a_1$ with 
a dipole solution outside. In Figure \ref{B_eos}, we plot the function 
$a_1(r)$ for different equations of state (EOS). The subscripts indicate the mass of 
the star. For example, WFF$_{14}$ corresponds to an equilibrium stellar model 
with EOS WFF (\cite{W1988}) and  $M=1.4 M_{\odot}$.  

The magnetic field $B^\alpha$ and the radial function $a_1$ are related by     
\begin{equation}\label{B_der}
B^\alpha=\epsilon^{\alpha \beta \mu \nu}u_\beta F_{\mu \nu} \, \, ,
\end{equation}
where $\epsilon^{\alpha \beta \mu \nu}$ is the Levi-Civita tensor, 
$u_\beta$ is the background four-velocity and $F_{\mu \nu}$ is the electromagnetic tensor.
Then equation (\ref{B_der}) leads to
\begin{equation}\label{B_exp}
H_r=\frac{a_1e^{\Lambda}} {\sqrt{\pi}r^2}\cos \theta 
\;\;\;\;\;H_\theta=\frac{a_1^{'}e^{-\Lambda}} {\sqrt{4\pi}}\sin \theta \, \, ,
\end{equation}
where we define $H_\alpha := B_\alpha/\sqrt{4\pi}$.
Note, however, that the coordinates defined by (\ref{new_trasf}) are not 
the usual magnetic coordinates known in the literature, see for example 
\cite{All1996} for a review, since they mix the poloidal angle and the 
radial component of the magnetic field in $(X,Y)$.

When we  transform  equation (\ref{2d}) according to (\ref{new_trasf}), we get:
\begin{equation}\label{1d}
\begin{split}
&\pi r^4 A_{tt}\frac{\partial ^2 {\cal Y}}{\partial t^2}=\frac{1}{4}a_1 {a_1}'^{2}\frac{\partial^2 {\cal Y}}{\partial Y^2}+\frac{y}{2}a_1\left({a_1}{''}-\frac{{a_1}'^{2}}{2}\right) \frac{\partial {\cal Y}}{\partial Y}\\
&+Y\left\{\frac{a_1{'}}{2}(\Phi-\Lambda){'}Y^2-X^2\left[\left(2\pi j_1-\frac{a_1}{r^2}\right)e^{2\Lambda}+\frac{a_1{''}}{2}\right]\right\}\frac{\partial {\cal Y}}{\partial Y}\\
&+XY^2\left\{\frac{a_1{'}}{2}(\Phi-\Lambda){'} +\left[\left(2\pi j_1-\frac{a_1}{r^2}\right)e^{2\Lambda}-\frac{a_1{''}}{2}\right]\right\}
\frac{\partial {\cal Y}}{\partial X} \, \, .
\end{split}
\end{equation}
By substituting $a_1{''}$ from equation (\ref{gf})  into the 
second and third term on the right hand side of equation (\ref{1d}), we arrive at 
\begin{equation}\label{1d_n}
 A_{tt}\frac{\partial ^2 {\cal Y}}{\partial t^2}=\tilde{A}_{10}\frac{\partial^2 {\cal Y}}{\partial Y^2}+\tilde{A}_{01} \frac{\partial {\cal Y}}{\partial Y} \, \, ,
\end{equation}
where
\begin{eqnarray}\label{coeff_1d}
\tilde{A}_{10}&=&\frac{1}{4 \pi r^4}a_1 {a_1}'^{2} \\
 \tilde{A}_{01}&=&\frac{X}{2\pi r^4}a_1\left(\frac{2}{r^2}a_1 e^{2\Lambda}-4\pi j_1e^{2\Lambda}-\frac{{a_1}'^{2}}{2}\right) \, \, ,
\end{eqnarray}
i.e., the coefficient of $\partial  {\cal Y}/\partial X$ vanishes identically.
The equation (\ref{1d_n}) is a $1+1$-dimensional wave equation. The perturbations propagate 
only along the $Y$ direction. In the last direction, the problem reminds of a study of 
waves propagating along the strings of a musical instrument.
 
\begin{figure}
\begin{center}
\includegraphics[height=8cm,angle=-90]{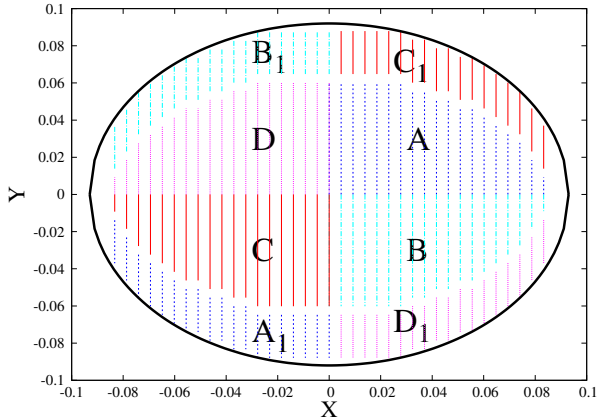}
\caption{
 Plot of the magnetic field lines in the $(X,Y)$ coordinates. Note that, after the point $X(a_{1\; \rm max})$ and $Y(a_{1\; \rm max})$,  lines of the same color continue 
in different quadrants. For example, the lines of the domain $A$ continue 
in the domain $A_1$. }%
\label{B_lines_XY}
\end{center}
\end{figure}

The construction of the grid in these $(X,Y)$ coordinates needs special
attention. The function $a_1$ has a maximum inside the star, and the
position of this maximum depends on the strength and
the topology of the magnetic field. A strong magnetic field, and also 
a toroidal field added to the standard poloidal field, pushes
the maximum near the surface of the star. 
The position of $a_{1\; \rm max}$ determines also the number of the closed magnetic field lines confined inside the star. Since the closed magnetic field lines don't reach the surface, they are subject to different boundary conditions than the open magnetic field lines. This 
difference leads to a different behaviour of the oscillations of open and closed 
magnetic lines, as will be shown later.  
In practice, we split the problem into two parts for both cases, i.e., for closed 
and open magnetic field lines. 
For the open magnetic field lines, the perturbation function
 $u(X,Y)$ is evolved until some points $X_{\rm max}$ and
$Y_{max}$ that correspond to the maximum of $a_{1\; \rm max}$, i.e., we work on 
the part $A$ of Figure \ref{B_lines_XY}.
Then the evolution continues, in the part $A_1$ of Figure \ref{B_lines_XY}, from 
the points $X_{\rm max}$ and
$Y_{max}$ and ends at the points $X_{\rm surf}$ and $Y_{\rm surf}$ that
correspond to the surface of the star. Note that in this second part,
we have to use the transformation 
\begin{equation}\label{transf_1}
X=-\sqrt{a_1} \sin \theta \;\;\; Y=-\sqrt{a_1} \cos \theta \, \, ,
\end{equation}
due to the sign change of the
derivative of $a_1$. In a similar fashion for the closed magnetic lines, we 
evolve $u(X,Y)$ until $X_{\rm max}$ and
$Y_{\rm max}$, but then the simulation doesn't end on the surface, but at $X_{\rm eq}$ and
$Y_{\rm eq}$, that correspond to points on the equator. From a technical point of 
view, we just ``{\it open}''
the field lines, then we solve the two problems separately and 
match the two solutions that we get via the boundary conditions. This procedure 
will be explained later. Figure \ref{B_lines_XY} may help in understanding 
the procedure followed in solving this problem. 
 The equation (\ref{1d_n}) must be solved with the appropriate boundary conditions. The 
boundary condition are different for open and closed magnetic lines. The open 
lines reach the surface and match up with the dipole magnetic field outside the star. 
Then, as we will show later, they couple on the surface and 
oscillate together. 
Differently, the closed magnetic lines confine themselves in 
the interior of the star, forming closed loops, and they never reach the surface. 
So they don't have any coupling to the open magnetic lines and 
oscillate independently. When we `open' the closed magnetic lines, creating magnetic 
strings, we open the closed loop, but don't change the boundary conditions 
at the ends of the strings.
 
In the $(X,Y)$ coordinate system, the boundary conditions become
\begin{itemize}
\item regularity at the center: ${\cal Y}(X,Y)=0$ for $X=0$ and $Y=0$,
\item no traction on the surface for the open lines:  
\begin{equation}\label{coupling}
\frac{a_1{'}}{2a_1}\left[X\frac{\partial {\cal Y}}{\partial X}+ Y\frac{\partial {\cal Y}}{\partial Y}\right]=0; 
\end{equation}
\item axisymmetry  at $X=0$:
\begin{equation}\label{coupling1}
Y\frac{\partial {\cal Y}}{\partial X}=0; 
\end{equation}
\item equatorial plane symmetry for $\ell=2$ initial data:  ${\cal Y}(X,Y)=0$ at $Y=0$. 

\item for $\ell=3$ initial data at $Y=0$:
\begin{equation}\label{coupling2}
Y\frac{\partial {\cal Y}}{\partial X}=0; 
\end{equation}
\end{itemize}

As a consequence, for initial data with $\ell=2$, the open `magnetic strings' are 
coupled on the surface of the star and 
near the axis for $X=0$ (corresponding to $\theta=0$). With initial data $\ell=3$,
there is an additional coupling at the equator (corresponding to $Y=0$).

\subsection{Numerical method}

\begin{figure}
\begin{center}
\includegraphics[height=8cm,angle=-90]{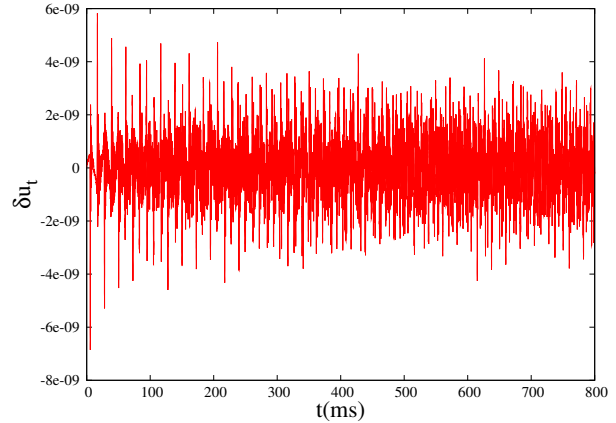}
\caption{
The time evolution of $\partial_t {\cal Y}$ for an open line in a point near the Y-axis  for the stellar equilibrium model WFF$_{14}$ and $\zeta=0$.
}%
\label{evol}
\end{center}
\end{figure}

We study a variety of equilibrium models, constructed for three representative EOSs
and different masses that range from $1.4-2.0$ $M_\odot$, see Table \ref{tab1}. We 
set the magnetic
field strength to $B_\mu=4\times
10^{15}$ Gauss, matching the field strength extrapolated from 
the analysis of the observations for
SGR $1806-20$ and
SGR $1900+14$. 
We construct a numerical equidistant grid $90 \times 90$ in the
$(X,Y)$ coordinates, setting $X_{\rm max}=\sqrt{a_{1\; \rm max}}\sin \theta$ 
and $Y_{\rm max}=\sqrt{a_{1\; \rm max}}\cos \theta$ and varying 
$\theta$ from $0$ and $\pi/2$. The structure of our grid 
is displayed in the first quadrant of Figure \ref{B_lines_XY}. Note 
that each  `magnetic string' in the first quadrant continues in the 
third quadrant. 
We evolve these two parts in time like two 
separate problems, taking into account 
the different boundary conditions for open and closed magnetic field lines. 
 The two parts are linked in every time step, requiring the continuity 
of the function ${\cal Y}$ and its 
derivative ${\cal Y}{'}$ . For our evolution, we use an iterative Crank-Nicholson scheme and
we evolve the perturbations for more than $1$s (see Figure \ref{evol}). In order to check 
the stability of the scheme,  we performed
an evolution for $2$s and found that our scheme was still
stable. 

We evolve the perturbations by providing initial data corresponding to either odd, 
$\ell=3$ or even, $\ell=2$, boundary conditions at $Y=0$. A change in the value 
of $\ell$ leads to different boundary conditions for open and closed lines and leads to different scaling in the oscillation frequencies of the open field lines.

We compute the FFT of various `strings' and find that each
`string' corresponds to a different frequency. 
 In \cite{SKS2008}, the authors find a family of 
upper frequencies near the
magnetic axis for $\theta=0$, corresponding to points along the 
Y-axis in our coordinates, and a family of lower frequencies for 
$\theta=\pi/2$, corresponding to points along the X-axis in our coordinates.
They identify the fundamental frequency peaks of the two families of 
QPOs with the edges or turning points of the Alfv\'en continuum suggested 
by  \cite{Levin2007}. These two frequencies were named 
fundamental lower $L_0$, for $\theta=\pi/2$, and fundamental upper  
$U_0$, for $\theta=0$ QPOs. The overtones are indicated by 
$L_n$ and $U_n$. In the following, we adopt these conventions 
when we refer to the oscillations of the open field lines, while we call 
the frequencies of the closed field lines $C_n$.

\begin{figure}
\begin{center}
\includegraphics[height=8.0cm,angle=-90=3.8cm]{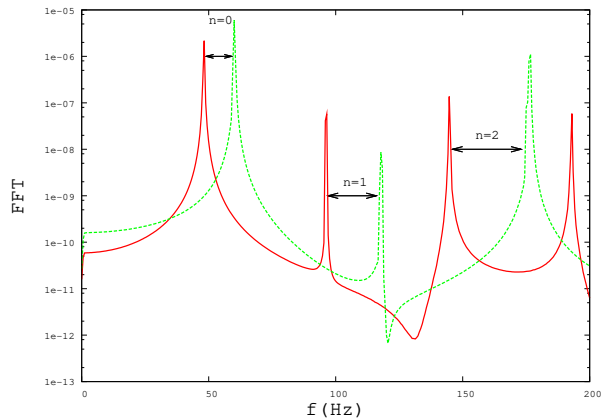}
\caption{
The FFT of  a point near the Y-axis, green/dashed line, and near the critical 
point, red/continuous line, for open lines and even parity, $\ell=2$.The stellar 
model is WFF$_{14}$.  }%
\label{fft}
\end{center}
\end{figure}

Our results confirm partially the results of \cite{SKS2008}. For open lines, a family 
of upper frequencies appears near the Y-axis, while a family of lower 
frequencies appears near the last open magnetic field line, just before the region of closed lines. 
This can be explained by the fact
that the `strings' near the Y-axis correspond to a stronger average magnetic field 
because in the $(r,\theta)$  coordinates they correspond to denser field 
lines, see Figure \ref{Opoint}. Since the frequency of the Alfv\'en oscillations 
is proportional to the strength of the magnetic field, we expect that these frequencies 
are the upper ones.  In addition, 
we find that, for all the 'magnetic strings', the upper
frequencies are multiples of the fundamental frequency both in the 
case of $L_n$ and in the case of $U_n$, see Figure \ref{fft}. However, 
this multiplicity is different when we consider even $\ell=2$ or odd $\ell=3$,
 initial data. In 
fact, we obtain 
\begin{eqnarray}
\label{rapp}
f_{L_n^{\rm even}} & \simeq & (2n+1)f_{L_0}
\\
\label{rapp1}
f_{U_n^{\rm even}} &\simeq & (2n+1)f_{U_0}
\end{eqnarray}
\begin{eqnarray}
\label{rapp2}
f_{L_n^{\rm odd}} & \simeq & (n+1)f_{L_0}
\\
\label{rapp3}
f_{U_n^{\rm odd}} &\simeq & (n+1)f_{U_0} \, \, , 
\end{eqnarray}
where the superscripts {\em even} and {\em odd} indicate results derived for even
$\ell=2$ and odd $\ell=3$ parity initial data and boundary conditions on the equator.
The different scaling shown in the relations (\ref{rapp}) - (\ref{rapp1}) 
and (\ref{rapp2})-(\ref{rapp3}) is due to
the different boundary conditions used for odd and even values of $\ell$. 

The problem  
can be understood by looking at a simple 
problem of two strings of finite length connected by 
by another 
spring. If the two strings are connected at just one of their ends, 
while the others are kept fixed, like in the case of even parity
where the 
 `magnetic strings' are coupled just near the surface 
and are fixed at the equator,  
then we obtain the relations  (\ref{rapp}) and  (\ref{rapp1}). 
This type of configuration was analyzed in \cite{K1986},  
in order to explain the w-modes of neutron stars. 
An analytic 
derivation of the relations (\ref{rapp}) and (\ref{rapp1}) can be found there.
If instead the two strings are coupled in both ends, like in the case of odd parity, 
then the relations (\ref{rapp2}) - (\ref{rapp3}) follow.

\begin{figure}
\begin{center}
\includegraphics[height=8.0cm,angle=-90=3.8cm]{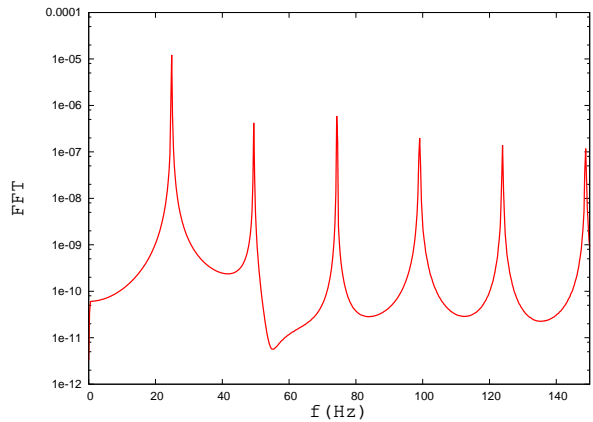}
\caption{
The FFT of the first closed line for the stellar equilibrium model WFF$_{14}$. It 
is worth noticing the spacing of the peaks in accordance to 
(\ref{rapp_c})}%
\label{fft_c}
\end{center}
\end{figure}

The frequencies of the closed magnetic field lines show a different behaviour. 
The upper and lower frequencies are scaling as, see also Figure \ref{fft_c},
\begin{equation}
\label{rapp_c}
f_{C_n}  \simeq  (n+1)f_{C_0} \, \, .
\end{equation}

The difference between the relation (\ref{rapp}) and (\ref{rapp_c}) is due 
to the different boundary conditions. The closed lines never reach the surface. They 
start from the equator where they close. This means that they are subject to the same 
boundary condition at both of their ends which resembles  
the toy problem of 
a string that oscillates with both ends fixed and gives the $n+1$ behaviour, 
as in the case odd values of $\ell$ for the open field  lines.
The frequencies that we find for closed lines initially decrease 
to a minimum and from then on increase. 
This behaviour results from two factors. First, from the fact that closed
lines are shorter than open lines, and, second, from the fact that 
the magnetic field becomes somehow locally weaker as we move off the axis. This 
behaviour of the spectrum has been identified after comparing our data with the 
results coming from the non-linear evolutions presented in \cite{DSF2009}.

The relations here provide the limits of the continuous part of the spectrum, in
accordance with the results of \cite{Levin2007}.   
Our results are summarized in Table \ref{tab1}. 


\subsection{Why we don't need artificial viscosity}

\begin{figure}
\begin{center}
\includegraphics[height=8cm,angle=-90]{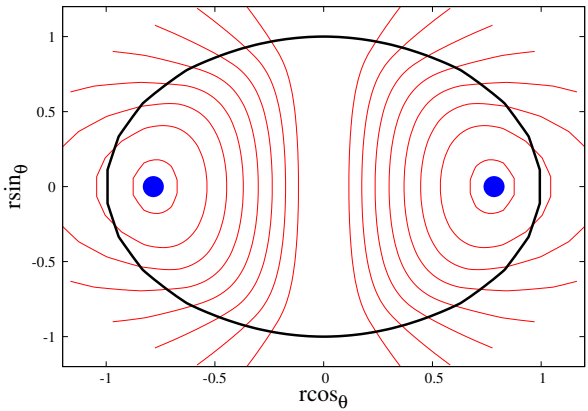}
\caption{
The magnetic field lines for the equilibrium stellar model WFF$_{14}$ and $\zeta=0$.
}%
\label{Opoint}
\end{center}
\end{figure}

In the preceding work \cite{SKS2008}, an artificial viscosity was added
in order to avoid a fast growing instability. This instability is due to the presence
of the so called ``\emph{O-type} neutral point'' (\cite{T1979}). If
the magnetic field lines of a poloidal magnetic field are closed and 
confined to the stellar interior,
then there is a point on which these magnetic field lines converge and become
degenerate (see Figure \ref{Opoint}). This point corresponds to 
the maximum of the function $a_1$. 

In our scheme, the magnetic field lines are practically straight, 
see Figure \ref{B_lines_XY}, even in the vicinity 
of the point that corresponds to $a_{1\;\rm max}$. 
This behavior is due to our choice to `open' the grid and to evolve the open 
and closed field lines as two independent problems. 
When we look at the point 
$a_{1\; \rm max}$, we find that it practically doesn't evolve and, because the 
`strings' are coupled only along the surface and near the $X=0$ axis, 
cannot be 
influenced by the oscillations of the other `strings'. In this way, we avoid the 
singular behaviour shown in the $(r,\theta)$ coordinates. The last is the main 
reason  
why our code is stable also without artificial viscosity.

\subsection{Toroidal fields}
 
\begin{figure}
\begin{center}
\includegraphics[height=8cm,angle=-90]{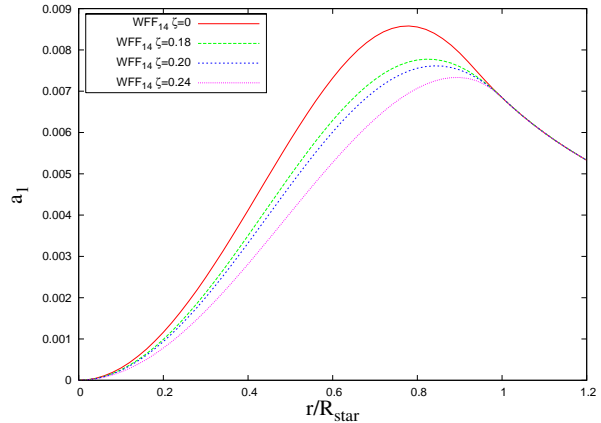}
\caption{
The function $a_1$ as a function of the normalized coordinates 
$r/R$ for an equilibrium model constructed with the equation of state WFF and 
a mass $M=1.4M_\odot$. Different curves reflect different 
values of the parameter $\zeta$, i.e., different strength of the 
toroidal magnetic field. 
}%
\label{B_wff}
\end{center}
\end{figure}

From equation (\ref{B_der}), the toroidal magnetic field can be written as
\begin{equation}\label{B_phi}
H_\phi=-\frac{\zeta e^{-\phi}a_1}{\sqrt{4\pi}}\sin \theta \, \, , 
\end{equation}
where $\zeta$ is a constant 
only depending on the toroidal magnetic field. By increasing 
or decreasing $\zeta$, we can influence the strength of the toroidal magnetic field. 
So $\zeta$ can be interpreted as a parameter that describes the strength of the 
toroidal magnetic field in relation to the poloidal field, see \cite{G2007} 
for details. 
The presence of a toroidal field modifies equation (\ref{gf}) 
by adding an extra  
term $\zeta^2e^{-2\nu}a_1$ (see \cite{SCK2008}).       
\begin{equation}\label{gf1}
a_1{''}e^{-2\Lambda}+\left(\Phi +\Lambda \right){'}e^{-2\Lambda}a_1{'}+\left(\zeta^2 e^{-2\Phi}-\frac{2}{r^2}\right)a_1=-4\pi j_1 \, \, .
\end{equation}
Hence, the toroidal field  influences our evolution equation (\ref{1d_n}) only  through
the presence of the extra term $a_1{''}$. 
By solving equation (\ref{gf1}) for 
$a_1{''}$ and subsequent substitution into equation (\ref{1d_n}), 
 the toroidal magnetic field  appears in the 1+1 equation (\ref{1d_n}) only through 
the parameter $\zeta$ in the term
\begin{equation}\label{c2}
\tilde{A}_{01}=\frac{X}{2}a_1\biggl(\frac{2 e^{2\Lambda}a_1}{r^2}-\zeta^2 a_1 e^{-2\Phi}-\frac{{a_1}'^{2}}{2}-4\pi j_1\biggr) \, \, .
\end{equation}
This means that the toroidal field affects the local propagation speed 
of the Alfv\'en modes. Also, the toroidal field pushes the position 
of $a_{1 \; \rm max}$ outwards and contributes to the stability of the 
magnetic field in the star. In Figure \ref{B_wff}, we plot the function $a_1$ 
for the EOS WFF$_{14}$ and for different values of the parameter $\zeta$. It is clear 
that when $\zeta$ increases, i.e., the strength of the toroidal magnetic field 
increases, then $a_{1\; \rm max}$ is pushed closer to the surface .

Hence, it is natural to expect changes in the oscillation spectrum  
due to the presence of the toroidal component of the magnetic field. We find that 
the frequencies are lower compared to the case in which 
only a poloidal field is present, see Table \ref{tab2}, especially 
for the closed magnetic `strings'. { \it This can be explained because, by 
the presence of a toroidal magnetic field $a_{1 \;\rm max}$, 
i.e., the point were 
the magnetic field lines converge, moves outwards, and hence the `magnetic strings' 
increase in length.}   
Since the frequency is inversely proportional to the length of the `strings', 
as the length increases, the frequency decreases.

\begin{table*}
  \centering
  \caption{Frequencies of Alfv\'en QPOs for a 
representative sample of equilibrium models, constructed for different EOS 
and fixed magnetic field strength $B=4 \times 10^{15}$ Gauss. The frequencies 
are computed for points near the  critical point, $L_n$, and near the 
Y-axis, $U_n$, for open lines, both for even and odd parity, and for the closed 
line ($C_n$).   }\label{tab1}
\vskip 24pt
\begin{tabular}{@{}lrrrrrrrr@{}}
\hline
Model & $M/R$ & $n$ &$f_{L_n^{\rm even}}(Hz)$ & $f_{L_n^{\rm odd}}(Hz)$ & $f_{U_n^{\rm even}}(Hz)$&$f_{U_n^{\rm odd}}(Hz)$& $ f_{C_{n}}(Hz)$\\      
\hline
        &    &      0& 23.00  & 46.56   & 28.05& 58.90 & 23.56  \\
$WFF_{14}$ & 0.189& 1& 69.55 & 93.11  & 84.15&118.4 &   47.68  \\
         &         & 2& 116.1& 139.1  &144.3 & 178.5 & 70.11 \\ 
\hline
       &    &      0& 13.46  & 27.48    & 16.27& 33.09 &  14.02 \\
$WFF_{18}$& 0.264& 1& 39.82& 54.41 & 48.24&66.75 &    28.05  \\
         &         & 2& 67.31& 81.89   &80.21 &99.84  &  42.06 \\ 

\hline
      &    &      0& 23.01  &47.06   & 28.24& 61.18 & 24.04  \\
$APR_{14}$& 0.178& 1&70.60 & 95.17 & 86.28& 121.8&   74.78 \\
         &         & 2& 117.1& 142.2  &144.3 & 182.5&  180.0 \\ 

\hline
      &    &      0& 19.04  &39.68   & 21.69& 44.97 & 19.57  \\
$APR_{16}$& 0.206& 1&58.72 & 79.88 & 66.6& 89.40&   39.68 \\
         &         & 2& 96.28& 119.6  &109.5 & 133.8&  60.31 \\ 
\hline
      &    &      0& 12.53 & 25.06  & 15.25& 31.60  & 13.07\\
$APR_{20}$& 0.269& 1& 37.59&50.09  & 44.12& 61.56  & 25.60 \\
         &         & 2& 62.10& 74.63  &74.63 &92.61  & 38.13 \\

\hline
      &    &      0& 13.10  & 27.83  & 16.78& 35.61  & 13.52 \\
$L_{20}$&  0.198& 1& 40.11&55.66  &  51.16&70.40 &  27.85 \\
         &         & 2& 68.35&83.50   &85.54 &105.6 &  41.34  \\ 

\hline
\end{tabular}
\end{table*}

\begin{table*}
  \centering
  \caption{Frequencies of Alfv\'en QPOs and their ratios for different stellar models with a toroidal component of the magnetic field and a surface magnetic field $B=4 \times 10^{15}$ Gauss. For open magnetic `strings' the frequencies are taken for  points near 
the critical point, $L_n$, and near the Y-axis, $U_n$, both for even and 
odd parity. We also give the value of the frequencies for the lower 
closed magnetic `string' ($C_n$). }\label{tab2}
\vskip 24pt
\begin{tabular}{@{}lrrrrrrrr@{}}
\hline
Model & $\zeta$ $(km^{-1})$ & $n$ &$f_{L_n^{\rm even}}(Hz)$ & $f_{L_n^{\rm odd}}(Hz)$ & $f_{U_n^{\rm even}}(Hz)$&$f_{U_n^{\rm odd}}(Hz)$& $ f_{C_{n}}(Hz)$\\      
\hline

        &    &      0& 15.15  & 30.30    &16.24&33.94  & 15.76  \\
$WFF_{14}$ & 0.24& 1& 46.06&  61.20  & 50.91&66.66&   31.15  \\
         &         & 2& 75.75& 90.90  &  81.20& 99.99&  79.38  \\ 
\hline
        &    &      0& 18.95  & 38.46    &22.85&49.6  & 19.51  \\
$APR_{14}$ & 0.18& 1& 56.83&  76.91  & 69.67&98.65&   39.01  \\
         &         & 2& 95.30& 114.8  &  117.0& 147.7&  57.96  \\ 
\hline
        &    &      0& 16.21  & 32.42    &19.35&41.84  & 16.73  \\
$APR_{14}$ & 0.20& 1& 48.11&  64.84  & 59.05&83.67&   33.47  \\
         &         & 2& 64.84& 97.79  &  99.88& 125.5&  51.25  \\ 
\hline
        &    &      0& 11.47  & 22.93    &13.92&29.90  & 11.88  \\
$L_{14}$ & 0.20& 1& 34.40&  45.46  & 42.18&58.57&   23.75  \\
         &         & 2& 56.52& 68.39  &  70.05& 88.46&  35.63  \\ 
\hline
\end{tabular}
\end{table*}

\section{Identification of the QPOs frequencies}

In a paper by \cite{W2006}, the timing analysis of the SGR 1806-20 and the SGR 1900+14 
has been computed. The study shows the appearance of several 
QPOs in the tail of these events. In particular, for
SGR 1806-20, the identified frequencies are $18,26,30,92,150,625$ and $1840$ Hz,  
while for SGR 1900+14, the frequencies are 28, 53, 84 and 155 Hz.  

Using the relations (\ref{rapp2}) and (\ref{rapp3}), we can easily show that 
for SGR 1900+14, if we consider $f_{L_0^{\rm odd}}=28$Hz, we can identify $f_{L_1^{\rm odd}}=56Hz$, 
and $f_{L_2^{\rm odd}}\simeq84$Hz.
In a similar way for SGR 1806-20, we have 
$f_{U_0^{\rm even}}=18$Hz and $f_{U_0^{\rm odd}}=30$Hz, and consequently 
$f_{U_2^{\rm even}}\simeq 92$Hz or $f_{U_2^{\rm odd}}\simeq 92$ , 
$f_{U_4^{\rm even}}\simeq 150$Hz or  $f_{U_4^{\rm odd}}\simeq 150$.
If we look at our results in the Tables 1 and 2, we can identify the 
QPOs in SGR 1900+14 with the lower frequencies for odd parity of a model 
with equation of state WFF and a mass around 1.8$M_ \odot$. For the 
QPOs from SGR 1806-20, we can fit the  observational data  by a model  
with equation of state APR, mass around 1.4$M_\odot$ and a toroidal field a little stronger than the poloidal field , $\zeta=0.18$.

\section{Discussion and conclusion} 

We have investigated torsional Alfv\'en modes of relativistic stars with 
poloidal and toroidal magnetic field using a new coordinate system $(X,Y)$, 
where $X=\sqrt{a_1}\sin \theta$ and $Y=\sqrt{a_1}\cos \theta$. In this new 
coordinate system, the perturbed two dimensional equation for the toroidal 
displacement ${\cal Y}(t,r,\theta)$ in \cite{SKS2008} is reduced to a $1+1$-dimensional 
equation, and the perturbations, i.e., Alfv\'en waves, are propagating only along 
the Y-direction. By 
solving the problem with the appropriate boundary conditions for open and closed 
magnetic `strings', we find that 
each `magnetic string' corresponds to a family of frequencies.  The upper frequency 
belongs to the family of open magnetic `strings' near the Y-axis because 
the Alfv\'en frequency is proportional to the local strength of the magnetic field,
and since near the Y-axis,
 the magnetic field is stronger than in the rest of the star. The family 
corresponding to the lower frequency is the one  just before the 
\textit{O-point}, i.e., the point where the magnetic lines converge and 
degenerate into a point. We find that for each family of frequencies 
there is a relation between the overtone frequencies and the fundamental 
frequencies, and this relation changes if we consider odd or even parity, 
see equations (\ref{rapp}), (\ref{rapp1}), (\ref{rapp2}), (\ref{rapp3}). In 
addition, we find a third family of frequencies linked to the closed lines. 
This family shows a minimum, whose value is similar to that of the lower frequencies 
of the open lines for even parity. Also, we find that in presence of a 
toroidal magnetic field the frequencies 
are lower than in the case with only a dipole magnetic field, both for 
closed and open magnetic  `string'. This is due to the 
change in the position of $a_{1\; \rm max}$ that moves outwards. This results in  
longer `magnetic strings' and, as consequence, to a lower characteristic frequency
for each `string'.

In a future work, it will be interesting to study whether the presence of a 
solid crust can quantitatively and qualitatively alter our results.   

\section*{Acknowledgments}

This work has been done in parallel with the work \cite{DSF2009}, where the same 
problem has been studied by using non-linear evolutions. We benefited from exchange of 
ideas and results with the authors and are particularly thankful for collaboration. We 
are also grateful to Hajime Sotani for helpful comments and for providing us with 
details of his numerical code. This work was supported via the 
Transregio 7 ``Gravitational Wave Astronomy" financed by the 
Deutsche Forschungsgemeinschaft DFG (German Research Foundation).


\end{document}